# High-Performance Flexible All-Perovskite Tandem Solar Cells with Reduced $V_{OC}$-Deficit in Wide-Bandgap Subcell

Huagui Lai[1], Jincheng Luo[2], Yannick Zwirner[1], Selina Olthof[3], Alexander Wieczorek[4], Fangyuan Ye[5], Quentin Jeangros[6,7], Xinxing Yin[8], Fatima Akhundova[9], Tianshu Ma[10], Rui He[2], Radha K. Kothandaraman[1], Xinyu Chin[6,7], Evgeniia Gilshtein[1], André Müller[1], Changlei Wang[10], Jarla Thiesbrummel[11], Sebastian Siol[4], José Márquez Prieto[9,12], Thomas Unold[9], Martin Stolterfoht[5], Cong Chen[2, *], Ayodhya N. Tiwari[1], Dewei Zhao[2, *] and Fan Fu[1, *]

[1] Laboratory for Thin Films and Photovoltaics, Empa - Swiss Federal Laboratories for Materials Science and Technology, Duebendorf, Switzerland
[2] College of Materials Science and Engineering, Engineering Research Center of Alternative Energy Materials & Devices, Ministry of Education, Sichuan University, Chengdu, Sichuan, 610065 China
[3] Department of Chemistry, University of Cologne, Greinstrasse 4–6, 50939, Cologne, Germany
[4] Laboratory for Surface Science and Coating Technologies, Empa − Swiss Federal Laboratories for Materials Science and Technology, Switzerland
[5] Soft Matter Physics and Optoelectronics, University of Potsdam, Potsdam, Germany
[6] Photovoltaics and Thin Film Electronics Laboratory, Institute of Electrical and Microengineering, École Polytechnique Fédérale de Lausanne, Neuchâtel, Switzerland
[7] Sustainable Energy Center, Centre Suisse d'Electronique et de Microtechnique (CSEM), Jaquet-Droz 1, 2002 Neuchâtel, Switzerland
[8] China-Australia Institute for Advanced Materials and Manufacturing, Jiaxing University, Jiaxing, Zhejiang, 314001 China
[9] Department of Structure and Dynamics of Energy Materials, Helmholtz-Zentrum-Berlin, Berlin, Germany
[10] School of Optoelectronic Science and Engineering & Collaborative Innovation Center of Suzhou Nano Science and Technology, Key Lab of Advanced Optical Manufacturing Technologies of Jiangsu Province & Key Lab of Modern Optical Technologies of Education Ministry of China, Soochow University, Suzhou, 215006, China
[11] Clarendon Laboratory, University of Oxford, Parks Road, Oxford OX1 3PU, United Kingdom
[12] Institut für Physik and IRIS Adlershof, Humboldt-Universität zu Berlin, 12489 Berlin, Germany

* Corresponding author: chen.cong@scu.edu.cn, dewei_zhao@hotmail.com, fan.fu@empa.ch

# ABSTRACT

Among various types of perovskite-based tandem solar cells (TSCs), all-perovskite TSCs are of particular attractiveness for building- and vehicle-integrated photovoltaics, or space energy areas as they can be fabricated on flexible and lightweight substrates with a very high power-to-weight ratio. However, the efficiency of flexible all-perovskite tandems is lagging far behind their rigid counterparts primarily due to the challenges in developing efficient wide-bandgap (WBG) perovskite solar cells on the flexible substrates as well as the low open-circuit voltage ($V_{OC}$) in the WBG perovskite subcell. Here, we report that the use of self-assembled monolayers as hole-selective contact effectively suppresses the interfacial recombination and allows the subsequent uniform growth of a 1.77 eV WBG perovskite with superior optoelectronic quality. In addition, we employ a post-deposition treatment with 2-thiopheneethylammonium chloride to further suppress the bulk and interfacial recombination, boosting the $V_{OC}$ of the WBG top cell to 1.29 V. Based on this, we present the first proof-of-concept four-terminal all-perovskite flexible TSC with a PCE of 22.6%. When integrating into two-terminal flexible tandems, we achieved 23.8% flexible all-perovskite TSCs with a superior $V_{OC}$ of 2.1 V, which is on par with the $V_{OC}$ reported on the 28% all-perovskite tandems grown on the rigid substrate.

# INTRODUCTION

Metal halide perovskite materials have attracted enormous attention from both the academic and industrial communities due to their excellent optoelectronic properties. After a decade of intensive research, the best certified efficiency of a single-junction perovskite solar cell (PSC) has already reached 25.7%.[1] In addition, the broadly tunable bandgaps (~1.17-3.10 eV) of perovskites make them ideal materials for constructing tandem solar cells (TSCs), which is a feasible approach to exceed the Shockley-Queisser (SQ) limit for single-junction solar cells.[2] The past 5 years have witnessed a rapid advance in perovskite-based TSCs, surpassing the highest efficiency of single-junction building blocks.[3] For instance, monolithic perovskite/Si TSCs have achieved certified efficiency of 31.3%, and perovskite-based thin-film TSCs such as perovskite/organic, perovskite/copper indium gallium selenide (CIGS) and perovskite/perovskite (all-perovskite) TSCs also reached certified efficiencies of 23.4%, 24.2% and 28%, respectively.[1,4] Among these perovskite-based tandem technologies, all-perovskite TSCs are of particular interest as they could be deposited on flexible and lightweight substrates using low-temperature solution processing methods (such as slot-die coating, inkjet printing, etc.), which are compatible with high throughput roll-to-roll manufacturing, and thus promises very low manufacturing cost and low $CO_2$ footprint.[5]

However, there are two major challenges that need to be overcome to realize efficient all-perovskite solar cells on flexible substrates. First, the uniform deposition of functional layers onto the flexible substrates is more challenging than on the rigid glass due to the generally rougher surface and inferior mechanical robustness of flexible substrates. Up to now, there are only two reports on flexible all-perovskite TSCs which yielded 21.3% and 24.4% (certified) efficiency, respectively, much lower than the best value of an all-perovskite TSC (28%) reported on a rigid glass substrate.[4,6,7] Another major challenge is the notable loss in open-circuit voltage ($V_{OC}$) in the wide-bandgap (WBG) PSCs, which largely limits the progress of all-perovskite TSCs.[8] The large $V_{OC}$-deficit (defined as $E_g/q$-$V_{OC}$) in WBG PSCs is commonly attributed to the relatively low initial radiative efficiency of the cells due to the high defect densities within the perovskite absorber layer and at the perovskite/charge selective layer interface. Over the past years, many approaches, including compositional engineering,[9-11] additive engineering,[12-14] interfacial engineering,[15-17] and mixed-dimensional engineering,[18-21] have been explored and developed to reduce the $V_{OC}$-deficit to values as low as 0.41 V in PSCs with bandgap (~1.7 eV) suitable for perovskite-Si tandems. However, extending these strategies to mitigate the $V_{OC}$-deficit in perovskites with even wider bandgaps (~1.80 eV) that are required for two-terminal (2T) all-perovskite tandems remains very challenging. This manifests as the relatively large $V_{OC}$-deficit of ~550 mV in WBG perovskite subcell used in the recently reported 26.4% monolithic all-perovskite tandem solar cells.[22]

Here, we report a $V_{OC}$-deficit of merely 480 mV in near-infrared (NIR) transparent WBG PSCs grown on flexible polymer foil by simultaneously reducing the $V_{OC}$ losses in perovskite bulk and at charge selective layer/perovskite heterojunctions. By replacing the conventional hole transport layer (HTL) poly-triarylamine (PTAA) with [2-(9H-carbazol-9-yl)ethyl]phosphonic acid (2PACz),[23] the perovskite absorber deposited on a flexible polymer foil shows better optoelectronic quality and uniformity, translating to ~ 40 mV gain in $V_{OC}$ for flexible NIR-transparent PSCs. Further passivating the perovskite with a post-deposition treatment (PDT) with 2-thiopheneethylammonium chloride (TEACl), the $V_{OC}$ of the flexible devices was improved by ~100 mV. With optimized TEACl PDT, we achieved a maximum efficiency of over 15% with a $V_{OC}$ of 1.29 V for a flexible NIR-transparent WBG PSC with a 1.77 eV bandgap, corresponding to a low $V_{OC}$-deficit of 480 mV. To the best of our knowledge, this is the lowest $V_{OC}$-deficit that has been achieved with a bandgap of ~1.80 eV on both rigid and flexible substrates. Combined with a flexible narrow bandgap (NBG) PSC, we demonstrate a proof-of-concept four-terminal (4T) TSC with a best PCE of 22.6%. Furthermore, we also achieved an all-perovskite 2T TSC with a remarkable $V_{OC}$ of 2.1V and a PCE of 23.8%. We note that this is the highest $V_{OC}$ for 2T flexible all-perovskite TSCs reported so far, demonstrating a further step towards achieving high-performance lightweight tandem solar cells with a reduced $V_{OC}$-deficit.

RESULTS AND DISCUSSION

The reference WBG PSC in our study was based on PTAA as HTL on the rigid substrates as well as flexible substrates, however, the device is clearly limited by its low $V_{OC}$ of ~1.15 V. To mitigate the severe $V_{OC}$ losses, we first tried to improve the hole-selective contact. To this end, we employed the self-assembled monolayer 2PACz as HTL which has been successfully employed in several recent studies to reduce surface recombination on the rigid indium tin oxide (ITO) substrate.[23,24] A schematic of our final device stack is shown in **Figure 1A**, where the PDT of the perovskite surface with TEACl is illustrated as well. **Figure 1B** shows the photograph of the multilayer flexible NIR-transparent WBG PSC.

Consistent with previous studies, we found that 2PACz delivers a higher $V_{OC}$ (~40 mV) and has better overall device performance than PTAA on the rigid ITO patterned glass substrates (**Figure S1**). However, when we replace the glass substrates with ITO patterned polyethylene naphthalate (PEN) substrates, the devices using 2PACz as HTL show a large spread in photovoltaic performance (**Figure S2**). In contrast, for PTAA-based devices, no difference is observed in terms of the spread of the PV parameters, which indicates the uniform deposition of PTAA on both rigid and flexible substrates. This result suggests that the deposition of 2PACz on PEN/ITO is not as uniform as that on glass/ITO, which might be due to poor anchoring of 2PACz on the relatively rough PEN/ITO surface (**Figure S3**). To ensure complete coverage of 2PACz on the ITO surface, we modified the deposition protocol of 2PACz by resting the 2PACz solution on the substrates for 60 s before starting the spinning and repeated this process twice to ensure good uniformity. Quasi-Fermi level splitting (QFLS) maps calculated from photoluminescence quantum yield (PLQY) measurements for different stack layers (**Figure S4**), show that the introduction of 2PACz between ITO and perovskite leads to a higher QFLS value, which corresponds to a higher $V_{OC}$ potential and suggests that 2PACz monolayer is more suitable than PTAA for achieving a lower $V_{OC}$ loss in the WBG PSCs.[25]

We performed the cross-sectional high-angle annular dark-field scanning transmission electron microscopy (HAADF-STEM) imaging as well as the energy-dispersive X-ray spectroscopy (EDX) mapping. As shown in **Figure 1C**, based on these measurements, we can estimate the thickness of the spin-coated WBG perovskite, PCBM and ZnO to be around 500, 35 and 25 nm, respectively, while the 2PACz is not visible at this magnification. The dashed rectangular area highlights the uniform film formation of PCBM and ZnO layers, which can also be distinguished by the signals of C and Zn in the EDX mapping. Here, chloroform (CF) was used as the solvent for the PCBM instead of commonly employed chlorobenzene (CB), as we observed a non-uniform deposition of the PCBM layer with CB. The analogous HAADF-STEM image and EDX mapping results for the cross-section of the device based on CB-processed PCBM are shown in **Figure S5**. A thick PCBM layer was obtained with CF as a solvent while with CB-processed PCBM is mostly thinner with a poor coverage. A possible explanation could be that CF exhibits a much lower boiling point (61.2 °C) than that of the CB (131.0 °C), which contributes to much faster evaporation during the fast spin-coating and forms a thicker and smoother PCBM layer.[26] Importantly, the better PV performance in case of the CF-processed PCBM is consistent with the improved morphology of the PCBM layer (**Figure S6**).

After the optimization of both the HTL and the ETL, we now targeted a reduction of interfacial recombination losses between these transport layers and the perovskite by means of surface passivation. In this regard, two-dimensional (2D) perovskites are known to be very effective at reducing the interfacial recombination losses and to improve the performance of PSCs.[27] To form a 2D perovskite on top of our 1.77 eV perovskite, we implemented the molecule 2-thiopheneethylammonium chloride, which has been previously used to create a 2D perovskite on top of a 1.68 eV perovskite.[21] Briefly, the TEACl PDT is performed by directly spin-coating TEACl solution onto the perovskite film, followed by a short annealing to dry the film (100 °C for 3 min). To optimize the performance, we varied the concertation of the molecule in an isopropanol solution and found that the PV performance of the devices was strongly affected, especially the $V_{OC}$ and the short-circuit current density ($J_{SC}$). The box charts of the PV parameters are presented in **Figure S7**. As shown in **Figure S7A**, a ~100 mV $V_{OC}$ improvement is observed for all the post-treated devices, indicating that TEACl PDT, regardless of the concentration, can effectively mitigate certain non-radiative pathways. Moreover, the $J_{SC}$ of the device increased significantly with an increasing concentration of TEACl PDT of up to 0.5 mg mL$^{-1}$, and then gradually decreased at higher concentrations (>0.5 mg mL$^{-1}$, **Figure S7B**). This change of the $J_{SC}$ is confirmed by the external quantum efficiency (EQE) measurements, as shown in **Figure S8**.

Overall, 0.5 mg mL$^{-1}$ of TEACl PDT effectively increases the PCE of the flexible NIR-transparent WBG PSC, mainly due to the largely improved $V_{OC}$. As shown in **Figure 1D**, the best PCE from this group (hereafter referred to as PDT) is 15.1%, with a $V_{OC}$ of 1.29 V, a $J_{SC}$ of 15.0 mA cm$^{-2}$, and a fill factor (FF) of 77.9%, while the best device without TEACl PDT (hereafter referred to as reference) has an efficiency of 13.1%, with a $V_{OC}$ of 1.191 V, a $J_{SC}$ of 14.6 mA cm$^{-2}$, and a FF of 75.4%. The dark current density-voltage ($J$-$V$) characteristics are provided in **Figure S9**. The superior $V_{OC}$ of 1.29 V for the PDT device is the highest value reported so far for a perovskite bandgap around 1.80 eV. The steady-state power output at the maximum power point of the reference and PDT devices, shown in **Figure S10**, is consistent with the results from $J$-$V$ scans. With light-intensity dependent $V_{OC}$ measurements (**Figure S11**), we extracted a smaller diode ideality factor $n$=1.25 for the PDT device than that of the reference device ($n$=1.45), which suggests a better diode quality and less non-radiative recombination for the device after a TEACl PDT.[28] In addition to the enhanced PV performance, the optimized device also shows good NIR-transparency. **Figure 1E** shows the transmittance and reflectance spectra of the PDT device. The full device demonstrates a high transmittance in the range of 720~880 nm (>80%) and 880~1100 nm (>70%), suggesting its great potential in the application of TSCs. Compared with previously reported $V_{OC}$ values of WBG (~1.80 eV) PSCs used in all-perovskite TSCs, our flexible and NIR-transparent device delivers the highest $V_{OC}$ and the lowest $V_{OC}$-deficit as shown in **Figure 1F** and **Table S1**.

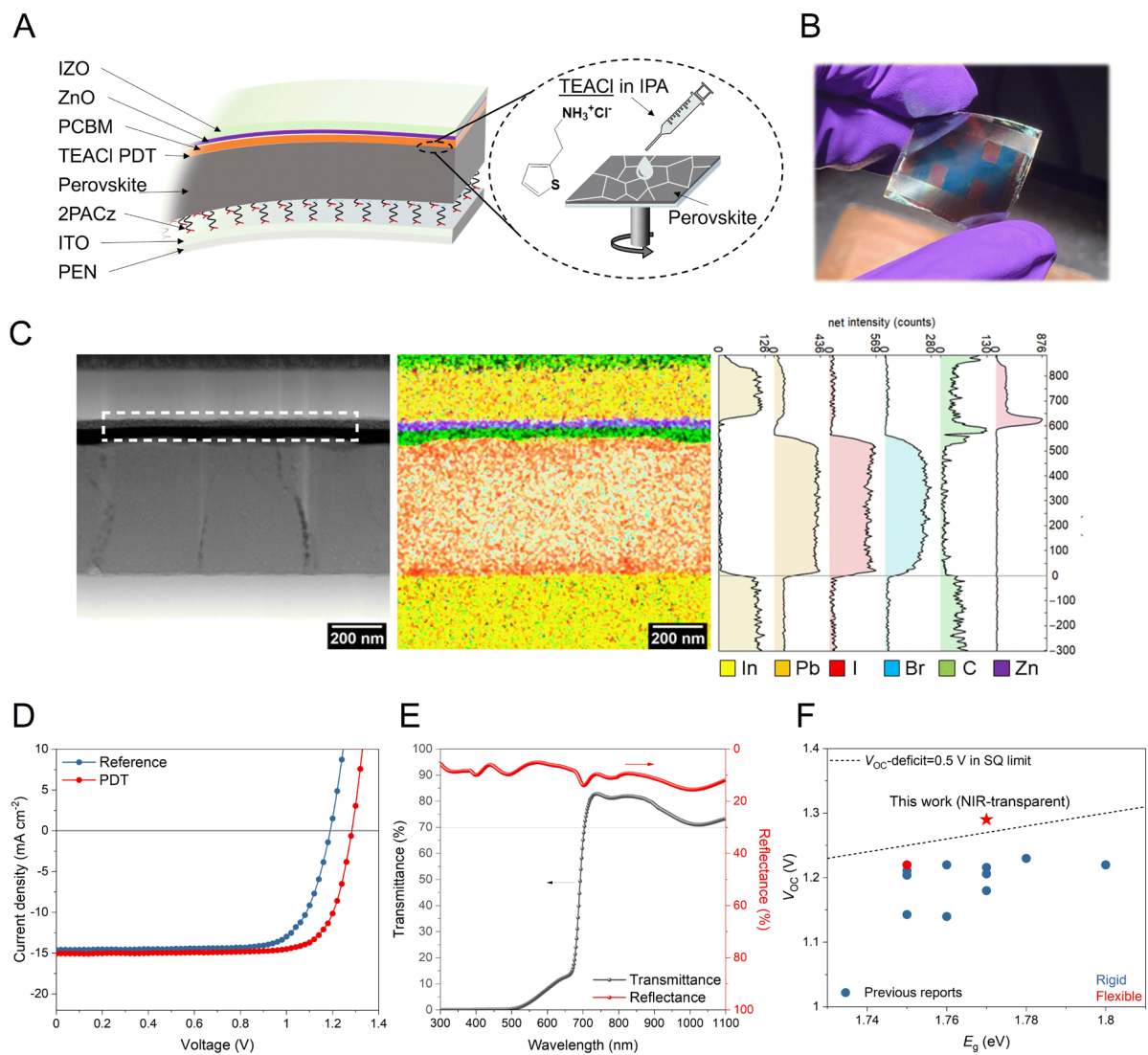

**Figure 1. Device architecture and photovoltaic performance of flexible NIR-transparent perovskite solar cells.**
(A) Schematic showing the configuration of the device, while the enlarged area shows the PDT of the perovskite surface by spin-coating TEACl in IPA.
(B) Digital photo of the flexible NIR-transparent wide-bandgap (1.77 eV) PSC.
(C) Cross-sectional HAADF STEM imaging and the corresponding energy-dispersive X-ray (EDX) mapping and line profiles for the device with 0.5 mg mL$^{-1}$ TEACl PDT.
(D) $J$–$V$ curves (reverse scans) of the best WBG PSCs with and without 0.5 mg mL$^{-1}$ TEACl PDT.
(E) Transmittance and reflectance spectra of the flexible NIR-transparent WBG PSC with 0.5 mg mL$^{-1}$ TEACl PDT.
(F) The achieved $V_{OC}$ as a function of bandgap extracted from all-perovskite tandem devices reported by recent studies. Detailed summary and references can be found in **Table S1**. The best $V_{OC}$ obtained in this work is shown for comparison. The SQ limit (maximum theoretical value) is indicated for comparison.

To obtain a more comprehensive understanding of TEACl PDT on the performance of our WBG PSCs, we performed a detailed study on the optoelectronic properties of the perovskite films and devices with various concentration of TEACl. From the absorption spectra and Tauc plots (**Figure S12**), an identical bandgap (~1.77 eV) was obtained from perovskite films with and without 0.5 mg mL$^{-1}$ TEACl PDT due to the negligible contribution of the surface layer. Scanning electron microscopy (SEM) images indicate a negligible effect of the PDT on the perovskite domain sizes. But at high TEACl concentrations (1.5 and 2 mg mL$^{-1}$), some black areas are observed on top of the perovskite films (**Figure S13**), which indicates that there are regions of different work function. These could be an accumulation of unreacted TEACl molecules on the surface. From atomic force microscopy (AFM) characterization, we found that the TEACl PDT tends to result in a rougher perovskite surface (**Figure S14**). The rougher perovskite surface could be a result of recrystallization upon the TEACl PDT. X-ray diffraction (XRD) patterns (**Figure S15**) show the formation of a diffraction peak at ~ 5.6° with an increased concentration of TEACl, accompanied by a gradually suppressed PbI$_2$ peak (12.6°) intensity. The emerging peak at 5.6° could be assigned to 2D phases TEA$_2$PbI$_4$, TEA$_2$PbI$_2$Br$_2$, or TEA$_2$PbI$_2$Cl$_2$ according to a previous report.[21] Therefore, we anticipate that a low dimensional phase is formed. However, based on XRD we are unable to identify which component(s) of the 3D perovskites is incorporated in that surface layer. Additional information is gained by analyzing the steady-state photoluminescence (PL) spectra of the perovskite films with TEACl PDT. We observe a PL emission peak at around 550 nm in the perovskite film with 3 mg mL$^{-1}$ TEACl PDT, suggesting that the 2D phase should be TEA$_2$PbI$_4$ (**Figure S16**).[21] It should be noted that the perovskite films with 0.5 or 2 mg mL$^{-1}$ TEACl PDT show no detectable PL peak at 550 nm, most likely due to the relatively low quantity of the 2D perovskite phase on the surface of the 3D perovskite.

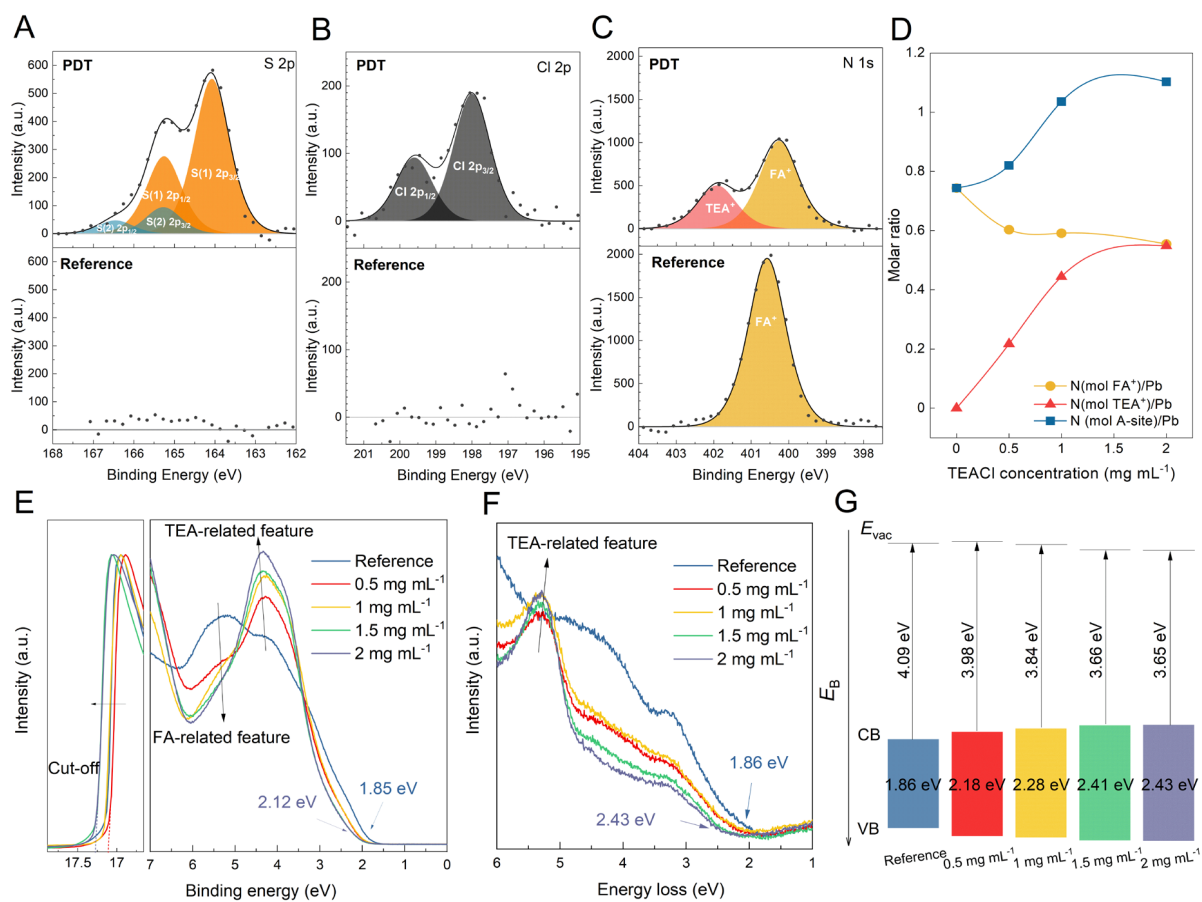

**Figure 2. XPS and UPS analysis of perovskite films.**
(A-C) XPS results for the reference and TEACl PDT perovskite films for the S 2p, Cl 2p and N 1s energy ranges, respectively. In C) the cation species are labelled based on the core level signals.
(D) Semi-quantitative molar ratios of $FA^+$ and $TEA^+$ on the surface relative to Pb as a function of TEACl concentration. The sum of both molar ratios is given by N (mol A-site). The herein presented values were calculated based on the atomic ratios from the N 1s and Pb 4f region taking the stoichiometry of the respective molecules into account. For both regions, the calculated information depth is 3 nm.
(E) UPS spectra of the perovskite films without (reference) and with 0.5, 1, 1.5, 2 mg mL$^{-1}$ TEACl PDT, respectively. The arrows indicate the onsets of the VBM.
(F) Reflection electron energy loss spectroscopy (REELS) spectra of the perovskite films without (reference) and with 0.5, 1, 1.5, 2 mg mL$^{-1}$ TEACl PDT, respectively. The arrows indicate the onsets, which corresponding to the bandgap energies.
(G) Schematic energy level alignment between perovskite films without (reference) and with 0.5, 1, 1.5, 2 mg mL$^{-1}$ TEACl PDT, respectively, extracted from the values determined in (E) and (F).

To elucidate the effect of the TEACl concentration on the perovskite's surface chemistry and electronic band structure, we performed X-ray photoelectron spectroscopy (XPS) measurements for the reference and TEACl PDT perovskite films. The S 2p and Cl 2p features were only found in perovskite films post-treated with TEACl, suggesting the presence of $TEA^+$ and $Cl^-$ on the perovskite surface. Interestingly, in addition to the main S 2p signal at 164 eV, a second small signal is observed at a higher binding energy. Similar observations have been previously reported for adsorbed thiophene derivatives on metal surfaces.[29,30] Consequently, it is possible that the additional strongly shifted sulfur core level signal originates from different environments of the $TEA^+$ molecules in contrast to $TEA^+$ molecules interacting with Pb atoms. Due to the expected loss in electron density at the sulfur, the feature at higher binding energies is assigned to the latter chemical state. Moreover, in **Figure 2C,** in case of the reference sample, we observe a single feature at ~401 eV for the reference perovskite film, similar to previous reports on the N 1s peak from $FA^+$.[31] With the TEACl PDT sample, an additional N 1s signal emerged at higher binding energies, which can be associated with $TEA^+$. Concurrently, the intensity for N 1s signal from $FA^+$ is significantly weakened. This could be due to a co-existence of $FA^+$ and $TEA^+$ as well as the formation of a $TEA^+$ rich surface layer, effectively attenuating the underlying $FA^+$ signal. Based on the atomic ratios of the N 1s features relative to the Pb 4f features (**Figure S17**), a semi-quantitative analysis of A-site cations relative to Pb located on the perovskite surface was performed. The results are summarized in **Figure 2D**, and indicate the gradual decrease of the $FA^+$ concentration with increasing concentration of TEACl PDT. Most notably, for TEACl concentrations higher than 0.5 mg mL$^{-1}$, an excess of A-site cations compared to Pb is observed. This is necessary for the formation of the proposed $TEA_2PbI_4$, and consistent with the emerging peak for the 2D phase in XRD patterns (**Figure S15**) above the same threshold.[32]

Ultraviolet photoelectron spectroscopy (UPS) and reflection electron energy loss spectroscopy (REELS) were used to investigate the effect of TEACl PDT on the surface electronic structures of the perovskite films. As shown in **Figure 2E**, with increased TEACl concentration, the FA-related density of states (DOS) feature is declining; while a new feature emerges that can be associated with the DOS of TEA. Similarly, in the REELS spectra in **Figure 2F**, a TEA-related feature emerges. The change in the observed DOS is consistent with the findings from XPS that the $TEA^+$ is likely accumulating at the surface and the signal from $FA^+$ is thus weakened. With the cut-off energies and onsets from UPS and REELS data, the energy level of the valence bands (VBs) of the perovskite films as well as the surface bandgap energies can be extracted, respectively, by using linear extrapolation. The determined energy level diagrams of the perovskite film surface are presented in **Figure 2G**. In particular, we observe a change in surface band gap by REELS from 1.86 eV (reference) to 2.43 eV (2 mg mL$^{-1}$ TEACl PDT), whereby the latter value is in good agreement with optical measurements for bulk $TEA_2PbI_4$, which has a reported bandgap of 2.33 eV.[21] The modification of the surface band structure of the perovskite with the 2D phase is likely beneficial to suppress the charge recombination at the interface between the perovskite and electron transport layer.

To gain a better understanding on whether the TEACl stays only on the perovskite film surface or there is any diffusion to the bulk, we performed XPS depth profiling for the perovskite film with the TEACl PDT. As evident from **Figure 3A**, the sulfur, which is exclusively found in the TEA$^+$ moiety, was only detected within the near surface area of the film and the signal quickly vanished with sputtering, indicating there is no TEA present in the bulk. In contrast to this, the Cl was detected not only at the perovskite surface but also found throughout the perovskite film. To confirm this observation, we further performed time-of-flight secondary ion mass spectroscopy (ToF-SIMS) depth profiling, which has a higher sensitivity than XPS for small amounts of different elements. As shown in **Figure 3B**, the S$^-$ signal shows again a sharp decay at the surface while Cl$^-$ has diffused into the bulk of the perovskite, fully consistent with the XPS depth profiling results (**Figure 3A & Figure S20**). We therefore can conclude that the TEA$^+$ moiety, represented by the S element, is confined to the surface where it likely forms a 2D perovskite phase on top of the perovskite film. While the majority of the Cl seems to stay on the surface as well, some diffuse throughout the perovskite film and might lead to an improvement of the quality of the bulk perovskite.[33,34]

Having clarified the effect of the TEACl treatment on the composition and electronic structure of the perovskite film, it is of interest to gain insight into the influence of TEACl PDT on the recombination kinetics of the perovskite films. Therefore, we measured the steady-state PL and time-resolved photoluminescence (TRPL) for the perovskite films with and without 0.5 mg mL$^{-1}$ TEACl PDT. As shown in **Figure 3C**, the perovskite film with PDT showed a much stronger PL emission than the reference one, manifesting an effective defect passivation effect with TEACl. The TRPL results in **Figure 3D** further confirmed this since the fitted effective carrier lifetime for the PDT film was 474 ns, which presents a significant increase compared to 278 ns for the reference film. This strongly indicates a reduced non-radiative recombination for the perovskite film with TEACl PDT. The improved carrier lifetime indicates that the TEA$^+$ moiety at the surface is likely to be effective in suppressing the defective states at the perovskite surface.[35] For Cl, both surface and bulk passivation are possible, considering the extended diffusion of Cl into the layer and its well-reported passivation effect.[36]

In addition, absolute PL measurements for different stack layers were carried out to break down the effect of the TEACl PDT on the $V_{OC}$ improvement. As shown in **Figure 3E**, we see that TEACl PDT improves the QFLS for all the stacks, both partial and full. The perovskite grown on the PEN foil exhibits a QFLS value of 1.34 V. After the TEACl PDT, the QFLS increases to 1.39 V, consistent with the suppression of nonradiative recombination on the perovskite surface. However, for the partial cell stacks with both the HTL (2PACz), and the ETL(PCBM) present, the QFLS values are much lower than the bare perovskite layer on glass, indicating that both the interfaces limit the $V_{OC}$, in particular, the perovskite/PCBM interface imposes the biggest QFLS deficit in both the reference and the PDT sample, which is also consistent with the fact that the QFLS of the *pin*-stack with both transport layers present is nearly identical to the QFLS of the perovskite/ETL stack. Clearly, the PDT is very effective at reducing the recombination induced by the ETL layer which leads to the enhanced $V_{OC}$.[37] Moreover, **Figure 3E** shows that there is a significant difference between the QFLS value of the reference *pin* sample and the $V_{OC}$ of the complete device. This behavior can for example be explained by an internal bending of the electron quasi-Fermi level, which affects primarily the $V_{OC}$ of the complete device rather than the QFLS in the perovskite layer. Considering that the internal QFLS value is significantly higher than the external $V_{OC}$ in the reference device, an energy misalignment between the perovskite and PCBM layer is suggested. With the application of the TEACl PDT, the QFLS-$V_{OC}$ difference is strongly reduced, which suggests an improved energy alignment between the perovskite and PCBM.[38]

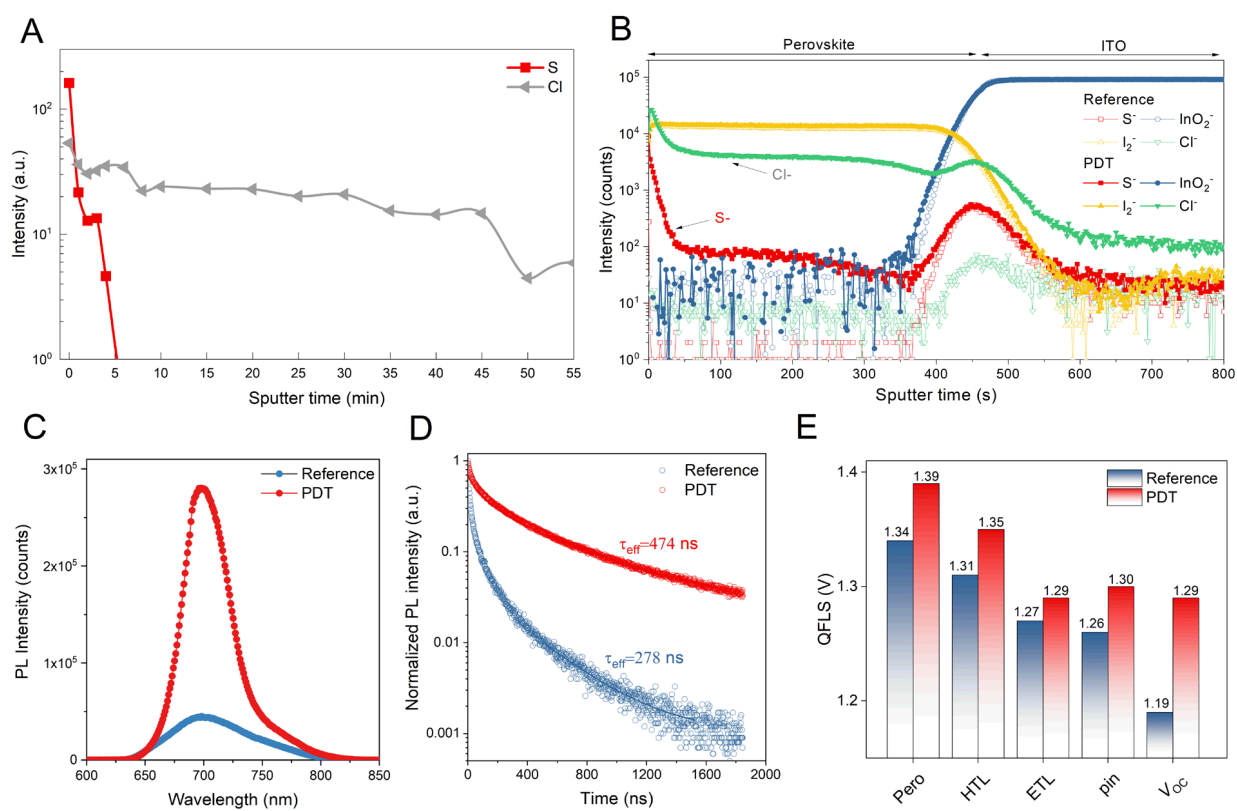

**Figure 3. Depth profiling of the elements and optoelectronic properties.**
(A) XPS depth profiles of S (Detailed spectra see **Figure S18**) and Cl (Detailed spectra see **Figure S19**) as determined in the perovskite film with TEACl PDT. These elements serve as markers for the TEA$^+$ and Cl$^-$ moiety of the TEACl reagent. The herein depicted total sputter time range reflects the depth profiling of the whole perovskite film (**Figure S20**).
(B) ToF-SIMS depth profiling for the perovskite films with and without TEACl PDT.
(C) PL results for the perovskite films with and without TEACl PDT at a 1 sun equivalent intensity at 350 nm excitation.
(D) TRPL decays for the perovskite films with and without TEACl PDT at a fluence of 1.65 nJ cm$^{-2}$ at an excitation wavelength of 375 nm. The fitted effective carrier lifetime is inserted close to the curves.

(E) QFLS values calculated from PLQY measurements for sample with different stack layers. Pero, HTL, ETL and *pin*-stack refer to neat perovskite, ITO/2PACz/perovskite, perovskite/PCBM, ITO/2PACz/perovskite/PCBM, respectively. $V_{OC}$ values are obtained from *J-V* measurements of the corresponding devices.

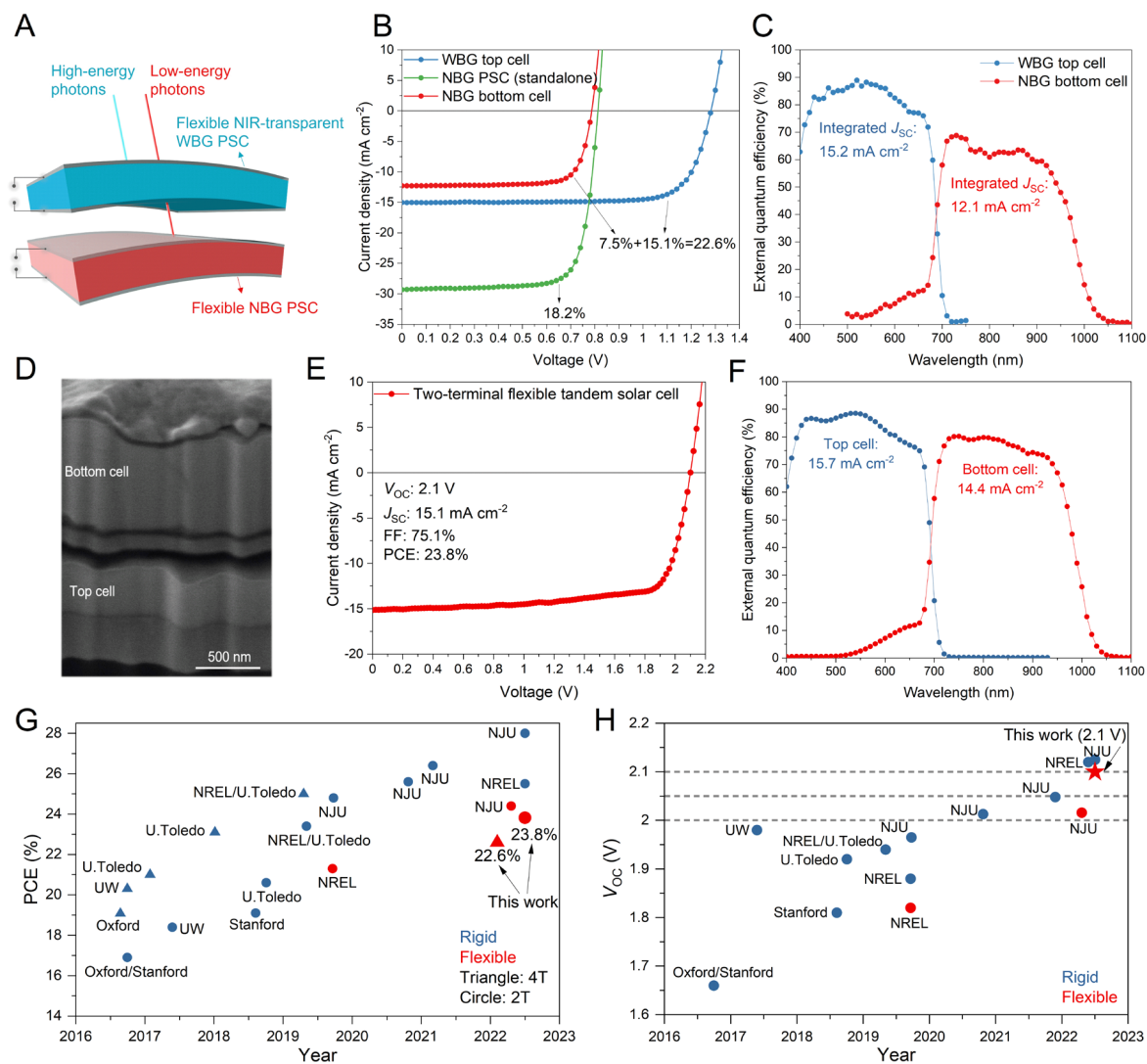

**Figure 4. Tandem architectures and best photovoltaic performance.**
(A) Schematics of the 4T all-perovskite flexible TSCs.
(B) *J-V* curves for 4T all-perovskite flexible TSCs.
(C) EQE spectra for 4T all-perovskite flexible TSCs.
(D) FIB-SEM image for 2T all-perovskite flexible TSC.
(E) *J-V* curve for the best-performing 2T all-perovskite flexible TSC.
(F) EQE spectra for 2T all-perovskite flexible TSC.
(G) Reported efficiencies progress for all-perovskite TSCs.
(H) Reported $V_{OC}$ values for 2T all-perovskite TSCs.

Finally, to demonstrate the potential of the flexible NIR-transparent WBG perovskite solar cells, we integrate them into flexible all-perovskite TSCs in both 4T and 2T configurations. **Figure 4A** shows the schematics of a 4T tandem device structure. NBG PSC used in 4T tandems is based on an architecture of PEN/ITO/PEDOT:PSS/(FASnI$_3$)$_{0.6}$(MAPbI$_3$)$_{0.4}$/C$_{60}$/BCP/Cu and it shows a PCE of 18.2%. An overall tandem PCE of 22.6% (15.1% from the top cell plus 7.5% from the filtered bottom cell) has been achieved. *J-V* curves are shown in **Figure 4B**, and detailed PV parameters are summarized in **Table S2**. The corresponding EQE spectra are also depicted in **Figure 4C** with integrated $J_{SC}$ values inserted. Notably, we report the best efficiencies for both the 1.24 eV and 1.77 eV bandgap single-junction flexible perovskite solar cells as compared in **Figure S21**. To go one step further, we also fabricated 2T all-perovskite flexible TSCs. **Figure 4D** shows a SEM image for the cross-section of our 2T flexible tandem device. The top cell and bottom cell are connected using atomic-layer-deposited $SnO_2$ and sputtered ITO. **Figure 4E** presents the *J-V* curve of our best-performing 2T flexible TSC (0.09 cm$^2$). Benefiting from the significantly suppressed $V_{OC}$-deficit with TEACl PDT for the WBG subcell, a high $V_{OC}$ of 2.1 V has been achieved in the 2T all-perovskite flexible TSC with a $J_{SC}$ of 15.1 mA cm$^{-2}$, FF of 75.1% and PCE of 23.8%. The steady-state power output of the 2T device shown in **Figure S22**, is consistent with the *J-V* efficiency. We have also cross-checked the *J-V* performance of the 2T device at EPFL PV-lab as shown in **Figure S23**. Notably, the $V_{OC}$ of 2.1 V of our flexible 2T TSC is higher than the $V_{OC}$ of the best performing rigid tandem (2.03 V) and flexible (2.0 V) tandem devices reported to date.[7,39] The corresponding EQE spectra for the 2T tandem device are measured and shown in **Figure 4F** with integrated $J_{SC}$ values of top and bottom cells inserted. The efficiency and $V_{OC}$ progress for all-perovskite TSCs are summarized in **Figure 4G** and **4H**, with detailed PV parameters summarized in **Table S3** and **S4**, respectively.

In conclusion, in this work we presented a multifaceted optimization approach of different device components. First, we used 2PACz as a hole transport layer for a 1.77 eV triple-cation perovskite device to effectively suppress the $V_{OC}$ loss at the HTL/perovskite interface and to enable the subsequent processing of high-quality uniform perovskite absorber layers on flexible ITO-patterned polymer foils. After that, we optimized the deposition of the PCBM transport layer by changing the solvent to ensure a good morphology. We then employed a TEACl PDT which forms a 2D perovskite on the perovskite surface to mitigate the decisive interfacial recombination loss at the electro-selective interface, optimize the band alignment between the perovskite/PCBM interface and lead to a reduced non-radiative recombination at the perovskite surface. Combining these optimization strategies, we achieved a high $V_{OC}$ of 1.29 V and 15.1%-efficient NIR-transparent WBG (1.77 eV) PSCs grown on flexible substrates. The high $V_{OC}$ corresponds to a record low $V_{OC}$-deficit of 480 mV for perovskite with a bandgap around 1.80 eV. In conjunction with flexible NBG (1.24 eV) PSCs, we achieved a PCE of 22.6% and 23.8% for flexible all-perovskite TSCs in 4T and 2T configurations, respectively. Moreover, we demonstrated the highest $V_{OC}$ of 2.1 V for flexible 2T all-perovskite tandem cells. As such, this work demonstrates how to achieve high-performance flexible TSCs by overcoming the large $V_{OC}$-deficit in WBG subcell through a sequential reduction of non-radiative bulk, surface and interface recombination losses. The high efficiency of our flexible tandem devices approaches to the best efficiencies for rigid ones and represents a significant step towards the commercialization of flexible and lightweight all-perovskite TSCs.

EXPERIMENTAL SECTION/METHODS

*Materials*: Prepatterned indium tin oxide (ITO) coated polyethylene naphthalate (PEN)(12 Ω/sq) were purchased from Advanced Election Technology Co., Ltd. Lead(II) iodide (PbI$_2$, 99.99%), cesium iodide (CsI, 99%), formamidinium iodide (FAI, ≥99.99%), formamidinium bromide (FABr, ≥99%), methylammonium bromide (MABr, ≥98%), methylammonium iodide (MAI, 98%), [2-(9H-Carbazol-9-yl)ethyl]phosphonic Acid (2PACz, >98%) were purchased from Tokyo Chemical Industry Co., Ltd. Dimethylformamide (DMF, anhydrous, 99.8%), dimethyl sulfoxide (DMSO, anhydrous, ≥99.9%), dimethyl ether (anhydrous, ≥99.9%), chlorobenzene (CB, anhydrous, 99.8%), chloroform (CF, anhydrous, 99.8%), isopropanol (IPA, anhydrous, ≥99.9%), lead(II) bromide (PbBr$_2$, 99.999%), tin(II) iodide (SnI$_2$, 99,99%), lead(II) thiocyanate (Pb(SCN)$_2$, 99.5%), copper (Cu, 99.99%) were purchased from Sigma-Aldrich Pty Ltd. Ethanol (anhydrous, ≥99.9%) was purchased from VWR International, LLC. Poly[bis(4-phenyl)(2,4,6-trimethylphenyl)amine], Poly(triaryl amine) (PTAA), [6,6]-Phenyl-C61-butyric acid methyl ester (PCBM), fullerene-C$_{60}$ and bathocuproine (BCP) were purchased from Xi'an Polymer Light Technology Corp. Zinc oxide nanoparticles (ZnO, 2.5 wt% in IPA) were purchased from Avantama AG. PEDOT:PSS (Clevios PVP Al 4083) was purchased from Heraeus Epurio LLC. All the materials were used as received. 2-thiopheneethylammonium chloride (TEACl) is synthesized according to a previous work.[21]

*Films preparation*: 1.2 M Cs$_{0.12}$FA$_{0.8}$MA$_{0.08}$PbI$_{1.8}$Br$_{1.2}$ Precursors were prepared by dissolving MABr (21.50 mg), CsI (74.83 mg), FABr (95.98 mg), FAI (198.11 mg), PbBr$_2$ (352.33 mg) and PbI$_2$ (663.85 mg) into a mixed solvent of DMF (1600 μL) and DMSO (400 μL). Before perovskite film spin-coating, the precursor solution was filtered with 0.22 um hydrophobic PTFE filters. The samples for top-view scanning electron microscopy (SEM), X-ray diffraction (XRD), atomic force microscopy (AFM), photoluminescence (PL) and time-resolved PL (TRPL), photoluminescence quantum yield (PLQY) measurements, X-ray photoelectron spectroscopy (XPS), ultraviolet photoelectron spectroscopy (UPS) and electron energy loss spectroscopy (EELS), time-of-flight secondary ionic mass spectrometry (ToF-SIMS) were performed with perovskite films on cleaned PEN/ITO substrates followed by different concentration of TEACl PDT. Samples for optical absorption spectrum measurements were performed with perovskite films and corresponding TEACl PDTs on white glasses.

*Device fabrication for WBG PSC*: Pre-patterned PEN/ITO substrates were first fixed onto rigid substrates with UV epoxy and then cleaned with ethanol and dried with $N_2$ flow. Before device fabrication, the substrates were further cleaned by UV/Ozone treatment (Jelight Company Inc.) for 30 min. 2PACz precursor (0.3 mg mL$^{-1}$ in ethanol, preheated at 55 °C) was spin-coated onto the cleaned ITO substrates at 3000 rpm for 30 s after 1 min's resting on the substrate. The same spin-coating step is performed twice to ensure a full coverage of 2PACz on the substrate, followed by an annealing at 100 °C for 5 min to remove the solvent. As a comparison, PTAA (5 mg mL$^{-1}$ in CB) was spin-coated onto the ITO substrate at 5000 rpm for 30 s, followed by an annealing at 100 °C for 5 min. After cooling, perovskite solution was spin-coated onto the substrate by a two-step spin-coating. The first step is 2000 rpm for 10 s with a ramp-up of 200 rpm s$^{-1}$. The second step is 6000 rpm for 40 s with a ramp-up of 2000 rpm s$^{-1}$. Diethyl ether (300 μL) was dropped onto the spinning substrate at the 20 s of the second step. The substrate was then annealed at 60 °C for 2 min and 100 °C for 7 min. TEACl PDTs were carried out by dissolving TEACl in IPA with varied concentration (0.5, 1, 1.5, 2 mg mL$^{-1}$) and spin-coating onto the perovskite film at 3000 rpm for 30 s, followed by an annealing at 100 °C for 3 min. After cooling, PCBM in CB or CF was spin-coated at 3000 rpm for 50 s, followed by annealing at 100 °C for 10 min. Thereafter, ZnO nanoparticles was spin-coated at 5000 rpm for 50 s, followed by annealing at 100 °C for 1 min. The substrates were then transferred to sputter chamber for the deposition of IZO electrode at a pulsed DC power of 200 W. For the comparison of 2PACz and PTAA as hole transport layer, a device configuration of glass/ITO/PTAA or 2PACz/Cs$_{0.12}$FA$_{0.8}$MA$_{0.08}$PbI$_{1.8}$Br$_{1.2}$/C$_{60}$/BCP/Cu was adopted. After the perovskite deposition, the samples were transferred into vacuum chamber for the thermal evaporation of C$_{60}$ (20 nm) and BCP (7 nm) layers, finished by evaporation of Cu (100 nm). For each substrate there are four devices. The designed contact area of each device is 0.1024 cm$^2$. The illuminated area of the device was defined with a patterned mask (0.09 cm$^2$). All spin-coating was carried out in $N_2$-filled glove box.

*Device fabrication for NBG PSC*: The pre-patterned PEN/ITO substrates were first fixed onto rigid substrates with UV epoxy and then cleaned with ethanol and dried with $N_2$ flow. Before device fabrication, the substrates were further cleaned by UV/Ozone treatment (Jelight

Company Inc.) for 30 min. PEDOT:PSS was spin-coated onto the cleaned substrate at 4000 rpm for 50s and then annealed at 120 °C for 20 min. Then the substrates were transferred into $N_2$-filled glove box. $(FASnI_3)_{0.6}(MAPbI_3)_{0.4}$ perovskite precursor was prepared as reported in previous work.[40] The $FASnI_3$ precursor solution was prepared by dissolving 372 mg $SnI_2$, 172 mg FAI, and 7.84 mg $SnF_2$ in 424 μL DMF and 212 μL DMSO. The $MAPbI_3$ precursor solution was prepared by dissolving 461 mg $PbI_2$ and 159 mg MAI, and 11.3 mg $Pb(SCN)_2$) in 630 μL DMF and 70 μL DMSO. Before mixing, the precursors were filtered with 0.22 um hydrophobic PTFE filters. Then stoichiometric amounts of $FASnI_3$ and $MAPbI_3$ perovskite precursors were mixed to obtain the $(FASnI_3)_{0.6}(MAPbI_3)_{0.4}$ precursor solution. The perovskite solution was spin-coated onto the substrate by a two-step spin-coating. The first step is 1000 rpm for 10 s with a ramp-up of 1000 rpm $s^{-1}$. The second step is 5000 rpm for 50 s with a ramp-up of 10000 rpm $s^{-1}$. Diethyl ether (700 μL) was dropped onto the spinning substrate at the 5 s of the second step. The substrate was then annealed at 65 °C for 3 min and 105 °C for 7 min. After the perovskite deposition, the samples were transferred into vacuum chamber for the thermal evaporation of $C_{60}$ (20 nm) and BCP (7 nm) layers, finished by evaporation of Cu (100 nm). The samples were then encapsulated with cavity glasses and UV epoxy.

*Device fabrication for 2T tandem*: All the procedures are the same for the WBG subcell until PCBM deposition is finished. The substrates were then transferred to ALD chamber for $SnO_2$ deposition (~ 20 nm) at relative low temperature (100 °C) by periodic pulse of tetrakis(dimethylamino) tin(IV) (99.9999%, Nanjing Ai Mou Yuan Scientific Equipment Co., Ltd) and deionized water. After ALD-$SnO_2$ deposition, the substrates were transferred to the magnetron sputtering system to sputter 100 nm ITO at a 30 W power under an Ar pressure of 0.4 Pa. Then, PEDOT:PSS diluted with IPA (volume ratio 1:1) was spin-coated onto the sputtered ITO substrates at 4000 rpm for 50 s and then annealed at 100 °C for 5 min in air. Next several steps to fabricate 2T tandem devices were consistent with those of fabrication of NBG PSCs.

*Film characterization*: The SEM images were taken with Hitachi S-4800 Scanning Electron Microscope using 5-10 kV acceleration voltage. The XRD patterns were measured on an X'Pert Pro (PANanalytical) in Bragg–Brentano geometry using Cu Kα1 radiation (λ = 1.5406 Å), scanning from 2 to 60 ° (2θ) with a step interval of 0.0167°. The AFM images were obtained using an AFM microscope (Bruker ICON3) in air. A silicon nitride tip (ScanAsyst-air) with a radius of 10 nm was used as the probe. The cantilevers' spring constant and resonant frequency were 0.4 N/m and 70 kHz, respectively. Steady-state photoluminescence (PL) and time-resolved photoluminescence (TRPL) were measured using FLS980 (Edinburgh Inc.). PL measurements were conducted using a 532-nm Xenon lamp with a monochromator while TRPL measurements were conducted using a 375-nm pico-second pulsed laser (EPL-375). Absorption of the films were obtained by measuring reflectance and transmittance using a Shimadzu UV/Vis 3600 spectrophotometer equipped with an integrating sphere. The reflectance data were corrected for the instrumental response stemming from diffuse and specular reflections both on the sample. Films thickness were measured by profilometer (AlphaStep P120).

*XPS measurement*: X-ray photoelectron spectroscopy was performed in a PHI Quantera system. Samples were analysed at a pressure of $10^{-9} - 10^{-8}$ Torr. The monochromatic Al Kα radiation was generated from an electron beam at a power of 12.6 W and a voltage of 15 kV. To minimize beam damage during measurements, the beam spot with a diameter of 50 μm was continuously scanned over an area of 500×1000 μm$^2$. Charge neutralization was performed using a low-energy electron source. Short-term measurements (<3 min) of the Pb 4f and C 1s core level before and after each presented measurement were conducted to rule out changes in the chemical state due to X-ray induced beam damage. The binding energy scale was reference to the main component of adventitious carbon at 284.8 eV, resulting in a typical inaccuracy of ±0.2 eV. Peak fitting of photoelectron features was performed in Casa XPS following Shirley-background subtraction using Voigt profile with GL ratios of 60. Atomic ratios were calculated using the instrument specific relative sensitivity factors. To estimate the information depth depending on the observed feature, the inelastic mean free path (IMFP) was calculated from the kinetic energy of the detected electrons based on the Tanuma, Powell, Penn formula.[41]

For depth profiling, Al Kα radiation generated by an electron beam at a power of 50.6 W and a voltage of 15 kV was utilized. The sample was etched using a beam of 2kV Ar$^+$ ions on an area of 2×2 mm$^2$ between each measurement.

*UPS measurement:* UPS measurements were performed using monochromatic UV source (VUV 5k, Scienta Omicron) at HeIα excitation (hv=21.22 eV) in combination with a hemispherical analyzer (Specs, Phoibis 100), set at a pass energy of 2eV. The samples were transferred under inert atmosphere and were at no point exposed to air.

*REELS measurement*: REELS measurements were performed with an electron excitation energy of 50 eV and a sample current of around 1μA, using a cold cathode (BaO based) electron gun (ELG-2, Kimball). The elastically and inelastically scattered electrons were measured with the same hemispherical analyzer as employed for the UPS measurement. The angle between gun and detector is 30° and the pass energy was set to 2eV.

*Absolute Photoluminescence Measurements*: Excitation for the PL measurements was performed with a 520 nm CW laser (Insaneware) through an optical fibre into an integrating sphere. The intensity of the laser was adjusted to a 1 sun equivalent intensity by illuminating a 1 cm$^2$ -size perovskite solar cell under short-circuit and matching the current density to the $J_{SC}$ under the sun simulator (~15 mA cm$^{-2}$ at 100 mWcm$^{-2}$, or 9.4x10$^{20}$ photons m$^{-2}$ s$^{-1}$). A second optical fiber was used from the output of the integrating sphere to an Andor SR393iB spectrometer equipped with a silicon CCD camera (DU420A-BR-DD, iDus). The system was calibrated by using a calibrated halogen lamp with specified spectral irradiance, which was shone into to the integrating sphere. A spectral correction factor was established to match the spectral output of the detector to the calibrated spectral irradiance of the lamp. The spectral photon density was obtained from the corrected detector signal (spectral irradiance) by division through the photon energy ($hf$), and the photon numbers of the excitation and emission obtained from numerical integration using Matlab. PL images for Figure S4 were recorded with Si CCD camera coupled to a liquid crystal tunable filter. The excitation source was 455 nm LED and the excitation photon flux was adjusted to 1.3 x 10$^{21}$ m$^{-2}$ s$^{-1}$. The system was calibrated to yield absolute photon flux using a calibrated halogen lamp source. QFLS provides an estimate for the maximum $V_{OC}$ that semiconductor absorber layer can achieve in a photovoltaic cell. QFLS distribution maps are calculated from the equation:

$$QFLS = qV_{OC}^{SQ} + K_b T ln(PLQY)$$

where $q$ is elementary charge, $V_{OC}^{SQ}$ is open-circuit voltage at Shockley-Queisser limit, $K_b$ is the Boltzmann constant, $T$ is temperature, PLQY is photoluminescence quantum yield.

The PLQY value is determined by the ratio of emitted photon flux to excited photon flux. The $V_{OC}^{SQ}$ is determined by approximating the bandgap from the measured PL peak emission energy.

*ToF-SIMS measurement:* Element depth profiles were obtained with a time-of-flight secondary ion mass spectrometer (ToF-SIMS V system, ION-TOF). The primary beam was 25 keV $Bi^{3+}$ with a total current of 0.38 pA and a raster size of 50 × 50 µm². $Cs^+$ ions were used with 1000 eV ion energy, 40 nA pulse current on a 400 × 400 µm² raster size to bombard and etch the film. The data were plotted with the intensity for each signal normalized to the total counts of the signal.

*STEM of WBG single junction cell:* The samples were prepared for TEM using the conventional lift-out method using a Zeiss NVision 40 dual beam FIB/SEM. The samples were then quickly transferred (< 2 min) to an FEI Tecnai Osiris microscope equipped with 4 silicon drift detectors for fast EDX mapping. STEM was performed at 200 kV with a beam current of 150 pA.

*FIB-SEM for 2T tandem solar cell*: The microstructure of the perovskite device stack was studied by scanning electron microscopy (SEM) (Helios NanoLab 600 DualBeam). The transfer into the FIB SEM was performed in air and the time in ambient air was <30 seconds. In a high vacuum (~10e-6 mbar), a cross-section was roughly milled with an ion beam current of 0.77 nA - followed by a cleaning cut with an ion beam current of 83 pA at 30 kV.

*Device characterization for single junction cell*: $J-V$ characteristics were measured in four-contact mode at standard test conditions (100 mW cm$^{-2}$) using a Keithley 2400 source meter. A solar simulator (ABA class, LOT-QuantumDesign) was calibrated to AM 1.5 G one sun illumination using a certified monocrystalline silicon solar cell (RS-ID-5, Fraunhofer-ISE, Ser. No. 114-2016). The solar cells were measured with an aperture mask with an active area of 0.09 cm² for each pixel. The $J–V$ measurements were performed in reverse direction (from $V_{OC}$ to $J_{SC}$, 100 mV/s) under 25 °C enabled by a cooling system. The steady-state efficiency as a function of time was recorded using a maximum power point (MPP) tracker, which adjusts the applied voltage to reach the maximum power point (perturb and observe algorithm). The external quantum efficiencies of the solar cells were measured with a lock-in amplifier. The probing beam was generated by a chopped white source (900 W, halogen lamp, 280 Hz) and a dual grating monochromator. The beam size was adjusted to ensure an illumination area within the cell area. The same single crystalline silicon solar cell as used in *J-V* characterization was used as a reference cell. White bias light was applied during the measurement with an intensity of ~ 0.1 sun. For the measurement of *J-V* and EQE characteristics of the filtered bottom cell, flexible NIR-transparent wide-bandgap top cell with a large active area (1 cm²) was used as a filter on top for easy cell alignment. Light-intensity dependent $V_{OC}$ characteristics were measured on the Paios measurement system (Fluxim AG).

*Device characterization for 2T tandem solar cell:* $J-V$ characteristics were measured using the same setup as that for single junction cell. The solar simulator spectrum was measured and compared with AM1.5G irradiation spectrum in **Figure S24**. The solar cells were measured with an aperture mask with an area of 0.09 cm$^2$ for each pixel in reverse direction. The steady-state efficiency as a function of time was recorded by fixing the bias voltage at $V_{pm}$, which is extracted from $J-V$ measurements. The cross-checking of the JV performance was done in EPFL PV-lab, in-house $J-V$ measurements were obtained using a temperature-controlled vacuum chuck at 25 °C, and a two-lamp (halogen and xenon) class AAA WACOM sun simulator with an AM 1.5 G irradiance spectrum at 1000 W/m$^2$. Independently certified SHJ cells were used to calibrate our solar simulator. Shadow masks were used to define the illuminated area (1.02 cm$^2$). The cells were measured with a scan rate of 100 mV/s (using an integration time of 0.1 s and a delay of 0.1 s for each data point, the voltage step was 0.02 V). The EQE characterizations of the 2T tandem solar cells were conducted in ambient air using an EQE setup (QE-R, Enlitech) in a near dark box. The monochromatic light ranging from 300 nm to 1100 nm was performed with a chopping frequency of 210 Hz, and the bias illumination from a 150 W white lamp was filtered with 550 and 850 nm optical filters for the measurement of bottom and top subcells' responses, respectively.

## ACKNOWLEDGEMENTS

This work was supported by funding from the European Union's Horizon 2020 research and innovation program under grant agreement no. 850937, the Strategic Focus Area Advanced Manufacturing under the project AMYS—Advancing manufacturability of hybrid organic–inorganic semiconductors for large-area optoelectronics, and Empa internal call 2021 (TexTandem). The work was financially supported by the National Key Research and Development Program of China (No. 2019YFE0120000), Fundamental Research Funds for the Central Universities (nos. YJ2021157 and YJ201955), Engineering Featured Team Fund of Sichuan University (2020SCUNG102). We acknowledge HyPerCells (a joint graduate school of the University of Potsdam and the Helmholtz-Zentrum Berlin) and the Deutsche Forschungsgemeinschaft (DFG, German Research Foundation) - project number 423749265 and 424709669 - SPP 2196 (SURPRISE and HIPSTER) for funding. We also acknowledge financial support by the Federal Ministry for Economic Affairs and Energy within the framework of the 7th Energy Research Programme (P3T-HOPE, 03EE1017C). H. L. thanks the funding of China Scholarship Council (CSC) from the Ministry of Education of P. R. China. S.O. acknowledges funding by the German Federal Ministry for Education and Research (MUJUPO$^2$, Grant OL 462/4-2). M.S. acknowledges the Heisenberg program from the Deutsche Forschungsgemeinschaft (DFG, German Research Foundation) for funding – project number 498155101. C. W. thanks the funding of National Natural Science Foundation of China (62005188), Natural Science Foundation of Jiangsu Province (BK20190825).

## AUTHOR CONTRIBUTIONS

F.F. conceived and directed the overall project. C.C., D.Z., A.N.T., F.F. designed and supervised this project. H.L. prepared the films, devices and characterizations on flexible WBG perovskite solar cells. H.L. and J.L. fabricated flexible 2T tandems. Y.Z. fabricated the NBG perovskite solar cell for flexible 4T tandems. S.O. conducted the UPS/REELS measurements and analysis. A.W. and S.S conducted the XPS measurements and analysis. F.Y., J.T., and M.S., performed absolute PL measurements and analysis. F.A., J.M.P and T.U. performed absolute PL measurements mapping and analysis. R.H. performed the PL/TRPL measurements. R.K.K. performed XRD and AFM measurements. E.G performed SEM measurements. Q.J. and A.M. conducted FIB-STEM and FIB-SEM measurements, respectively. X.Y. synthesized the TAECl. X.C cross-checked *J-V* performance of the 2T tandem. T.M. and C.W. performed EQE measurements for 2T tandems. H.L. and F.F. wrote the manuscript with input from all co-authors. All authors discussed the results and reviewed the manuscript.

## CONFLICT OF INTEREST

The authors declare no conflict of interest.